\documentclass{article}

\usepackage{arxiv}
\usepackage[utf8]{inputenc} 
\usepackage[T1]{fontenc}    
\usepackage{booktabs}       
\usepackage{amsfonts}       
\usepackage{nicefrac}       
\usepackage{microtype}      
\usepackage{lipsum}		
\usepackage{amsmath,amssymb}
\usepackage{graphicx}
\usepackage{color}

\newcommand{\grl}{    {Geophys. Res. Lett.}}
\newcommand{\jgr}{    {J. Geophys. Res.}}
\newcommand{\ssr}{    {Space Sci. Rev.}}

\newcommand{\aap}{    { Astronomy and Astrophysics}}
\newcommand{\solphys}{ {Solar Physics}}
\newcommand{\apj}{ {Astrophys. J. }}
\newcommand{\apjl}{    {Astrophys. J. Lett.}}

\newcommand{\mnras}{ {Mon.Not.Royal.Soc. }}

\newcommand{\apjs}{    {Astrophys. Journal. Suppl. Ser.}}

\newcommand{\blue}{\textcolor{black}}

\newcommand{\Zstroke}{%
  \text{\ooalign{\hidewidth\raisebox{0.2ex}{--}\hidewidth\cr$Z$\cr}}%
}

\def\Xint#1{\mathchoice
   {\XXint\displaystyle\textstyle{#1}}%
   {\XXint\textstyle\scriptstyle{#1}}%
   {\XXint\scriptstyle\scriptscriptstyle{#1}}%
   {\XXint\scriptscriptstyle\scriptscriptstyle{#1}}%
   \!\int}
\def\XXint#1#2#3{{\setbox0=\hbox{$#1{#2#3}{\int}$}
     \vcenter{\hbox{$#2#3$}}\kern-.5\wd0}}

\def\dashint{\Xint-}

\title{On quasi-parallel whistler waves in the solar wind}

\author{
    {Ivan Y. Vasko} \\
	Space Sciences Laboratory, University of California at Berkeley, California, CA94720, USA\\
	Space Research Institute of Russian Academy of Sciences, Moscow, 117997, Russia\\
	\texttt{ivan.vasko@ssl.berkeley.edu} \\
	\and
	{Ilya V. Kuzichev} \\
	New Jersey Institute of Technology, Newark, New Jersey, NJ07102, USA\\
	Space Research Institute of Russian Academy of Sciences, Moscow, 117997, Russia\\
	\and
	{Anton V. Artemyev} \\
	Institute of Geophysics and Planetary Sciences, University of California, Los Angeles, California, CA 90095, USA\\
	Space Research Institute of Russian Academy of Sciences, Moscow, 117997, Russia\\
	\and
	{Stuart D. Bale} \\
	Space Sciences Laboratory, University of California at Berkeley, California, CA94720, USA\\
	Department of Physics, University of California at Berkeley, California, CA94720, USA\\
    \and
	{John W. Bonnell} \\
	Space Sciences Laboratory, University of California at Berkeley, California, CA94720, USA\\
	\and
	{Forrest S. Mozer} \\
	Space Sciences Laboratory, University of California at Berkeley, California, CA94720, USA\\
}

\begin{document}
\maketitle

\begin{abstract}
	\blue{The recent simulations showed that the whistler heat flux instability, which presumably produces the most of quasi-parallel coherent whistler waves in the solar wind, is not efficient in regulating the electron heat conduction. In addition, recent spacecraft measurements indicated that some fraction of coherent whistler waves in the solar wind may propagate anti-parallel to the electron heat flux, being produced due to a perpendicular temperature anisotropy of suprathermal electrons.  We present analysis of properties of parallel and anti-parallel whistler waves unstable at electron heat fluxes and temperature anisotropies of suprathermal electrons typical of the pristine solar wind. Assuming the electron population consisting of counter-streaming dense thermal core and tenuous suprathermal halo populations, we perform a linear stability analysis to demonstrate that anti-parallel whistler waves are expected to have smaller frequencies, wave numbers and growth rates compared to parallel whistler waves. The stability analysis is performed over a wide range of parameters of core and halo electron populations. Using the quasi-linear scaling relation we show that anti-parallel whistler waves saturate at amplitudes of one order of magnitude smaller than parallel whistler waves, which is at about $10^{-3}\;B_0$ in the pristine solar wind. The analysis shows that the presence of anti-parallel whistler waves in the pristine solar wind is more likely to be obscured by turbulent magnetic field fluctuations, because of lower frequencies and smaller amplitudes compared to parallel whistler waves. The presented results will be also valuable for numerical simulations of the electron heat flux regulation in the solar wind.}
\end{abstract}

\keywords{solar wind, \and whistler waves \and heat conduction \and electron heat flux}

\section{Introduction \label{sec:intro}}

The spacecraft measurements showed that the electron heat conduction in the solar wind is less efficient than predicted by the collisional Spitzer-H\"arm law \cite{Ogilvie71,Montgomery72,Feldman75,Scime94,Bale13,Landi:2014a}. The measured electron heat flux is typically smaller than the Spitzer-H\"arm values and \blue{suppressed} below a threshold dependent on $\beta_{e}=8\pi n_{e}T_{e}/B^2$ \cite{Feldman76,Gary1999a,Gary1999b,Tong18,Tong19b:apj,Wilson19b}. \blue{Remote measurements indicated} that \blue{similar suppression of the electron heat conduction occurs} in the solar corona \blue{\cite{Wang15,Bian16:apj,Bian18:apj}} and various astrophysical environments \cite{Cowie77,Meiksin86,Zhuravleva19,Kunz19:arxiv}. The early spacecraft measurements around 1 AU showed that whistler waves might be a plausible wave activity regulating the electron heat conduction in collisionless or weakly-collisional solar wind plasma \cite{Feldman76,Feldman76b,Gary1999b}. \blue{However, until recently spacecraft measurements could not resolve} the occurrence and generation mechanisms of whistler waves in the solar wind.

\blue{The early spacecraft measurements of the electron velocity distribution function (VDF) in the solar wind indicated that whistler waves may be produced} by the whistler heat flux instability (WHFI) \cite{Gary75,Gary1994a,Abraham-Shrauner77}. The electron VDF in a slow solar wind around 1 AU can be often described by a combination of a dense thermal core and a tenuous suprathermal halo population \blue{\cite{Feldman75,Maksimovic97,Maksimovic05,Pierrard16,Lazar17:AA,Tong2019a}.} In the plasma rest frame, the core and halo populations are described by Maxwell and $\kappa-$distributions drifting in the opposite directions parallel to a local magnetic field. \blue{The net electron current is absent,} while the electron heat flux, carried predominantly by halo electrons, is parallel to the halo drift velocity. \blue{Gary et al. \cite{Gary75} showed that} at sufficiently large values of the halo drift velocity, the electron VDF is unstable to the WHFI and produces whistler waves propagating parallel to the electron heat flux \cite{Abraham-Shrauner77,Gary1994a,Shaaban18:mnras}. \blue{Until recently} the WHFI was considered to be the dominant mechanism producing whistler waves and \blue{regulating} the electron heat conduction in the solar wind \cite{Feldman76,Feldman76b,Gary1999b,Gary1994a}. \blue{The recent quasi-linear analysis \cite{Pistinner1998a,Shaaban19} and Particle-In-Cell (PIC) simulations \cite{Roberg-Clark:2016,Kuichev19,Lopez19} have conclusively shown that the WHFI cannot efficiently regulate the electron heat conduction in the solar wind and stimulated theoretical analysis of other instabilities potentially regulating the electron heat flux \cite{Horaites18:stability,komarov_2018,Shaaban18:mnras,Shaaban18:pop,Vasko2019a,Verscharen19:apj,Roberg-Clark19} as well as experimental analysis of whistler waves in the solar wind using modern spacecraft measurements \cite{Lacombe14,Stansby16,Kajdic:2016a,Tong19b:apj,Tong2019a}.}

\blue{Modern spacecraft measurements in the solar wind at 1 AU have shown that coherent whistler waves propagate within 20$^{\circ}$ of a local quasi-static magnetic field \cite{Lacombe14,Stansby16,Kajdic:2016a,Tong19b:apj}. The occurrence rate of whistler waves is positively correlated with the macroscopic electron temperature anisotropy $T_{e\perp}/T_{e||}$ and varies from a few up to a few tens of percent \cite{Tong19b:apj}}. Coherent whistler waves are identified in the magnetic field spectra as local bumps superimposed on the spectrum of turbulent magnetic field fluctuations persistently present in the solar wind \cite{Lacombe14,Tong19b:apj}. Noteworthy that the occurrence of whistler waves \blue{can be higher}, because the presence of a local bump \blue{may be} obscured by turbulent magnetic field fluctuations. Simultaneous measurements of the electron VDF and \blue{several whistler waveforms presented by Tong et al. \cite{Tong2019a}} have demonstrated that the WHFI \blue{indeed operates in the solar wind} and, consistently, the whistler waves propagate parallel to the electron heat flux. The extensive statistical analysis \blue{by Tong et al. \cite{Tong19b:apj}} of whistler waves identified \blue{as local bumps} in the magnetic field spectra has shown that \blue{properties of whistler waves in the solar wind are consistent with the WHFI, though the propagation direction of the whistler waves} (parallel or anti-parallel to the electron heat flux) could not be determined. \blue{We cannot} rule out that some fraction of whistler waves observed in the solar wind propagates anti-parallel to the electron heat flux and \blue{is} produced by a mechanism different from the WHFI. \blue{Moreover,} recent spacecraft measurements of electron VDF \blue{indicate} that anti-parallel whistler waves may be indeed present in the solar wind \blue{\cite{Pierrard16,Lazar17:AA,Tong19b:apj,Wilson19b,Wilson19c}}.

In the original WHFI theory, the core and halo electron populations were assumed to be isotropic or have parallel temperature anisotropy \cite{Gary75,Abraham-Shrauner77}. In this case only whistler waves propagating parallel to the electron heat flux could be unstable. However, the recent analysis of the electron VDF showed that the halo population may have perpendicular temperature anisotropy, $1<T_{\perp h}/T_{||h}\lesssim 1.5$ \blue{ \cite{Pierrard16,Lazar17:AA,Wilson19b,Wilson19c}}, \blue{and the statistical analysis of Tong et al. \cite{Tong19b:apj} demonstrated that the occurrence rate of whistler waves is positively correlated with the macroscopic electron temperature anisotropy.} At a negligible heat flux, the electron VDF with the halo population satisfying $T_{h\perp}/T_{h||}>1$ is unstable to the classical whistler temperature anisotropy instability (WTAI) \cite{Sagdeev60,Kennel66,GaryWang96}, which produces identical whistler waves propagating parallel and anti-parallel to \blue{a local} magnetic field. The presence of a non-negligible electron heat flux is expected to break that symmetry, resulting in different frequencies, growth rates and saturated amplitudes of parallel and anti-parallel whistler waves. The analysis of properties and effects of unstable anti-parallel whistler waves is currently of interest from experimental and theoretical perspectives, because parallel whistler waves turned out to be inefficient in regulating the electron heat conduction in the solar wind \blue{\cite{Pistinner1998a,Roberg-Clark:2016,Kuichev19,Shaaban19,Lopez19}}, while anti-parallel whistler waves may be efficient in scattering suprathermal electrons \cite{Vocks05,Pierrard11,Vocks12} and, hence, may be potentially efficient in the electron heat flux regulation.

In this paper, we present \blue{theoretical analysis} of properties of parallel and anti-parallel whistler waves that \blue{can} be unstable in the solar wind. \blue{The focus is on whistler wave parameters, which quantitative estimates are valuable} for spacecraft data analysis and numerical simulations of the electron heat flux regulation in the solar wind. In Section \ref{sec1}, we present basic formulas of the linear stability theory and typical parameters of core and halo electron populations in the solar wind around 1 AU. In Section \ref{sec2}, we address effects of power-law index of the halo $\kappa-$distribution on the WHFI growth rates. In Section \ref{sec3}, we compare properties of unstable parallel and anti-parallel whistler waves \blue{as well as expected saturated amplitudes estimated using the quasi-linear scaling relation inferred by Tao et al. \cite{Tao17} and Kuzichev et al. \cite{Kuichev19}}. In Section \ref{sec4}, we discuss valuable implications of the presented results.


\section{General formulas and typical parameters\label{sec1}}

Motivated by spacecraft measurements in the solar wind at 1 AU \cite{Lacombe14,Stansby16,Kajdic:2016a,Tong2019a,Tong19b:apj}, we restrict the linear stability analysis to whistler waves propagating parallel and anti-parallel to the electron heat flux which is, in turn, parallel to a uniform quasi-static magnetic field ${\bf B}_0$. The dispersion relation of parallel and anti-parallel whistler waves is given as follows \cite{Mikhailovskii74,Cuperman81}
\begin{eqnarray*}
  \frac{k^2c^2}{\omega^2}&=&1+\frac{\omega_{pe}^2}{\omega}\int \frac{d^3{\bf v}}{\omega-k v_{||}-\omega_{ce}}\frac{v_{\perp}^2}{2}\left[\frac{k}{\omega}\frac{\partial F_{e}}{\partial v_{||}}+\frac{\omega-k v_{||}}{\omega v_{\perp}}\frac{\partial F_{e}}{\partial v_{\perp}}\right]+\\&+&\frac{\omega_{pi}^2}{\omega}\int \frac{d^3{\bf v}}{\omega-k v_{||}+\omega_{ci}}\frac{v_{\perp}^2}{2}\left[\frac{k}{\omega}\frac{\partial F_{i}}{\partial v_{||}}+\frac{\omega-kv_{||}}{\omega v_{\perp}}\frac{\partial F_{i}}{\partial v_{\perp}}\right]\nonumber
  \label{eq:1}
\end{eqnarray*}
where $F_{e}(v_{||},v_{\perp})$ and $F_{i}(v_{||},v_{\perp})$ are electron and proton VDFs, $v_{||}$ and $v_{\perp}$ are velocities parallel and perpendicular to ${\bf B}_0$, $\omega=\omega_{r}+i\gamma$ is the whistler wave frequency, $k$ is the wave vector component along the magnetic field, $\omega_{pe}=(4\pi n_0e^2/m_{e})^{1/2}$ and $\omega_{pi}=(4\pi n_0e^2/m_{i})^{1/2}$ are electron and proton plasma frequencies, $\omega_{ce}=eB_0/m_{e}c$ and $\omega_{ci}=eB_0/m_i c$ are electron and proton cyclotron frequencies. Under typical conditions in the solar wind, whistler waves \blue{driven by resonant electrons} have frequencies \blue{$\omega\gg \beta_{i}\;\omega_{ci}$} and do not interact resonantly with thermal protons, because $(\omega+\omega_{ci}) \gg k (2T_{i}/m_i)^{1/2}$, \blue{where $\beta_{i}=8\pi n_0T_{i}/B_0^2$ is the proton beta typically of the order of one \cite{Wilson18:apj,verscharen19:review}}, $(\omega+\omega_{ci})/k$ is the cyclotron resonance velocity, and  $(2T_{i}/m_i)^{1/2}$ is the proton thermal velocity. Therefore, the contribution of the proton population, that is the third term on the right-hand side of the dispersion relation, can be replaced by \blue{$-\omega_{pi}^2/\omega(\omega+\omega_{ci})$}, which corresponds to \blue{approximation of cold protons} \cite{Mikhailovskii74}. In addition, we neglect the displacement current effects, that is unity on the right-hand side, because $kc/\omega\gg 1$, and rewrite the dispersion relation as follows
\begin{eqnarray}
  \frac{k^2c^2}{\omega^2}+\blue{\frac{\omega_{pi}^2}{\omega(\omega+\omega_{ci})}}\approx \frac{\omega_{pe}^2}{\omega}\int \frac{d^3{\bf v}}{\omega-k v_{||}-\omega_{ce}}\frac{v_{\perp}^2}{2}\left[\frac{k}{\omega}\frac{\partial F_{e}}{\partial v_{||}}+\frac{\omega-k v_{||}}{\omega v_{\perp}}\frac{\partial F_{e}}{\partial v_{\perp}}\right]
  \label{eq:2}
\end{eqnarray}
The singular denominator $(\omega-kv_{||}-\omega_{ce})^{-1}$ indicates that whistler waves can be unstable only due to interaction with electrons in the first normal cyclotron resonance, $v_{||}=(\omega-\omega_{ce})/k$ (see, e.g., Refs. \cite{Mikhailovskii74,Shklyar09}). 

The electron VDF is a combination of the core and halo populations, $F_e(v_{||}, v_{\perp})=f_{c}(v_{||},v_{\perp})+f_{h}(v_{||},v_{\perp})$. In the plasma rest frame, the VDF of the core population is a drifting Maxwell distribution
\begin{eqnarray*}
f_{c}(v_{||},v_{\perp})&=&\frac{n_{c}}{A_{c}}\left(\frac{m_{e}}{2\;\pi \;T_{c}}\right)^{3/2}\exp\left[-\frac{m_{e}(v_{||}-u_{c})^2}{2\;T_{c}}-\frac{m_{e}v_{\perp}^2}{2\;A_{c}\;T_{c}}\right]
\label{eq:fc}
\end{eqnarray*}
while the VDF of the halo population is a drifting $\kappa-$distribution (see, e.g., Ref. \cite{Pierrard10} for review on $\kappa-$distributions)
\begin{eqnarray*}
f_{h}(v_{||},v_{\perp})&=&\frac{n_{h}}{A_{h}}\left(\frac{m_{e}}{2\pi\;(\kappa-3/2)\;T_{h}}\right)^{3/2}\frac{\Gamma(\kappa+1)}{\Gamma(\kappa-1/2)}\left[1+\frac{m_{e}(v_{||}-u_h)^2}{(2\kappa-3)\;T_{h}}+\frac{m_{e}v_{\perp}^2}{(2\kappa-3)\;A_{h}\;T_{h}}\right]^{-(\kappa+1)}
\label{eq:fh}
\end{eqnarray*}
where $n_{\alpha}$, $u_{\alpha}$ and $T_{\alpha}$ are core and halo densities, parallel drift velocities and electron temperatures, $A_{\alpha}$ determines temperature anisotropies (subscript $\alpha=c,h$ corresponds to core and halo populations), $\Gamma(\kappa)$ is the Gamma function. A perpendicular temperature anisotropy corresponds to $A_{\alpha}>1$, while a parallel temperature anisotropy corresponds to $A_{\alpha}<1$. The electron current in the plasma rest frame is assumed to be zero, $n_{c}u_{c}+n_{h}u_{h}=0$. We assume the halo drift velocity to be parallel to the magnetic field, so that the electron heat flux is also parallel to the magnetic field and given as follows \cite{Gary1999b}
\begin{eqnarray}
q_{e}=\int v_{\parallel}\;\frac{m_{e}{\bf v}^2}{2}F_{e}({\bf v})\;d^3{\bf v}=\frac{1}{2}\sum_{\alpha=c,h} n_{\alpha} u_{\alpha} \left(3\;T_{\alpha}+2\;A_{\alpha}\;T_{\alpha} +m_{e}u_{\alpha}^2\right)
\label{eq:qe}
\end{eqnarray}
The electron heat flux is often normalized to the free-streaming heat flux value, $q_{0}=1.5 \;n_0\;T_{e}\;(2T_{e}/m_{e})^{1/2}$, where $n_0=n_c+n_h$ is the total electron density, $T_{e}=(n_cT_c+n_hT_h)/n_0$ is the total parallel electron temperature \cite{Cowie77,Gary1999b}.

For the electron VDF specified above the dispersion relation (\ref{eq:2}) can be written as follows 
\begin{eqnarray}
  \frac{k^2c^2}{\omega^2}+\blue{\frac{\omega_{pi}^2}{\omega(\omega+\omega_{ci})}}\approx \frac{\omega_{pe}^2}{\omega^2}\sum_{\alpha=c,h}\frac{n_{\alpha}}{n_0} \left[\frac{\omega-A_{\alpha}\;k\;u_{\alpha}}{kv_{\alpha}}{Z}_{\alpha}(\xi_{\alpha})+(A_{\alpha}-1)+(A_{\alpha}-1)\frac{\omega-\omega_{ce}}{kv_{\alpha}}{Z}_{\alpha}(\xi_{\alpha})\right]
  \label{eq:DispRel}
\end{eqnarray}
where $v_{\alpha}=(T_{\alpha}/m_{e})^{1/2}$ are core and halo thermal velocities, $\xi_{\alpha}=(\omega-ku_{\alpha}-\omega_{ce})\;/\;kv_{\alpha}$ is the argument of the plasma dispersion function ${Z}_{\alpha}(\xi)=\int_{-\infty}^{+\infty}dx\;z_{\alpha}(x)\;/\; (x-\xi)$, where 
\begin{eqnarray}
  z_{c}(x)=(2\pi)^{-1/2}\exp(-x^2/2),\;\;\; z_{h}(x)=\frac{\Gamma(\kappa)}{(2\pi)^{1/2}(\kappa-3/2)^{1/2}\Gamma(\kappa-1/2)}\left[1+\frac{x^2}{2\kappa-3}\right]^{-\kappa}
  \label{eq:Zaa}
\end{eqnarray}
The general expressions for the dispersion function corresponding to the $\kappa-$distribution \blue{can be found in Refs. \cite{Thorne&Sumers91,Mace95,Lazar08:pop,Lazar13:AA}}, while we assume whistler waves to be slowly growing, $\gamma\ll \omega_{r}$, which can be verified {\it a posteriori}, and use the standard asymptotic expansion \cite{Mikhailovskii74}
\begin{eqnarray*}
{Z}_{\alpha}(\xi)=\Zstroke_{\alpha}(\xi)-i\;\pi\;{\rm sign}(k)\;z_{\alpha}(\xi),\;\;\;\;\;
\Zstroke_{\alpha}(\xi)=\dashint_{-\infty}^{+\infty}dx\;\frac{z_{\alpha}(x)}{x-\xi}
\end{eqnarray*}
where $k>0$ and $k<0$ for parallel and anti-parallel whistler waves, respectively. The computed dispersion functions allow solving the dispersion relation (\ref{eq:DispRel}) using the standard method for slowly growing plasma modes \cite{Mikhailovskii74}. The whistler wave dispersion relation $\omega_r(k)$ is determined by solving $\Lambda(\omega_r,k)=0$, where
\begin{eqnarray}
    \Lambda(\omega,k)=&&\frac{k^2c^2}{\omega^2}+\blue{\frac{\omega_{pi}^2}{\omega(\omega+\omega_{ci})}}-\nonumber\\&-&\frac{\omega_{pe}^2}{\omega^2}\sum_{\alpha=c,h}\frac{n_\alpha}{n_0}\left[\frac{\omega-A_{\alpha}ku_{\alpha}}{kv_{\alpha}}\Zstroke_{\alpha}(\xi_{\alpha})+(A_{\alpha}-1)+(A_\alpha-1)\frac{\omega-\omega_{ce}}{kv_{\alpha}}\Zstroke_\alpha(\xi_\alpha)\right],
     \label{eq:omegar}
\end{eqnarray}
\blue{In the limit of cold and non-drifting electrons the solution $\omega_{r}(k)$ of equation $\Lambda(\omega_{r},k)=0$ should be $\omega_{r}^2(k)\approx\left[\omega_{ce} k^2c^2/(k^2c^2+\omega_{pe}^2)\right]^2+\omega_{ci}\omega_{ce} k^2c^2/(k^2c^2+\omega_{pe}^2)$, which is the dispersion relation of whistler waves in a cold plasma provided that $\omega_{pe}\gg \omega_{ce}$ (see Refs. \cite{Mikhailovskii74} and \cite{Shklyar04}). At $\omega\gg \omega_{ci}$ the cold dispersion relation of whistler waves is $\omega_{r}(k)\approx \omega_{ce} k^2c^2/(k^2c^2+\omega_{pe}^2)$.} The growth rate is computed as
\begin{eqnarray}
  \gamma(k)=-\frac{\pi\;{\rm sign}(k)}{\left(\partial \Lambda/\partial\omega\right)_{\omega=\omega_r(k)}}\sum_{\alpha=c,h}\frac{\omega_{pe}^2}{\omega_r^2}\frac{n_{\alpha}}{n_0} z_{\alpha}(\xi_{\alpha})\left[\frac{\omega_r-A_{\alpha} ku_{\alpha}}{kv_{\alpha}}+(A_{\alpha}-1)\frac{\omega_r-\omega_{ce}}{kv_{\alpha}}\right]
  \label{eq:gamma}
\end{eqnarray}
where $\xi_{\alpha}=(\omega_r-ku_{\alpha}-\omega_{ce})\;/\;kv_{\alpha}$.  

The normalization $\omega\rightarrow \omega/\omega_{ce}$ and $k\rightarrow kc/\omega_{pe}$ in the dispersion relation (\ref{eq:DispRel}) shows that the growth rate $\gamma/\omega_{ce}$ depends on $k c/\omega_{pe}$ and the following parameters of the core and halo populations
\begin{itemize}
    \item $n_{c}/n_0$: density of the core population with respect to the total electron density.
    \item $u_{c}/v_{A}$: core parallel drift velocity with respect to the Alfv\'{e}n speed, $v_{A}=B_0/(4\pi n_0 m_{i})^{1/2}$.
    \item $\beta_{c}=8 \pi n_{c}T_{c}/B_0^2$: core parallel beta parameter.
    \item $T_{h}/T_{c}$: halo to core parallel temperature ratio.
    \item $A_{c}$ and $A_{h}$: core and halo temperature anisotropies.
\end{itemize}
\blue{In what follows we assume the core population to be isotropic, $A_{c}=1$, because the perpendicular temperature anisotropy of the core population in the solar wind is typically small (see, e.g., Ref. \cite{Wilson19b} and Table \ref{tab:first_table}) and does not significantly affect the growth rate of whistler waves unstable due to perpendicular temperature anisotropy of halo electrons.} The effect of core and halo densities on the growth rate is twofold. First, the growth rate is proportional to the halo density, $\gamma/\omega_{ce}\propto n_{h}/n_0$, because the whistler waves are mostly resonant with halo electrons. Second, the growth rate is dependent on core and halo densities through the halo drift velocity, $u_{h}=-(n_{c}/n_{h})\;u_{c}$, but that dependence can be addressed by considering various core drift velocities $u_{c}$ rather than various $n_{c}/n_{h}$. In what follows we present the growth rates computed for $n_{c}/n_0=0.95$, while the growth rates at any other values of the core and halo densities can be estimated by adjusting the core drift velocity and taking into account that $\gamma/\omega_{ce}\propto n_{h}/n_0$. We note that the dispersion relation $\omega_{r}(k)$ and growth rate $\gamma(k)$ are independent of $\omega_{pe}/\omega_{ce}$, because we neglected the displacement current in Eq. (\ref{eq:2}). That is justified at $\omega_{pe}/\omega_{ce}\gg 1$, which is typical of the solar wind, where $\omega_{pe}/\omega_{ce}\sim 100$ \blue{\cite{verscharen19:review}}. 

The typical values of the critical parameters in the solar wind at radial distances from 0.3 to 4 AU are given in Table \ref{tab:first_table} along with parameters to be used in the stability analysis in Sections \ref{sec2} and \ref{sec3}. In the next section we consider $q_{e}/q_{0}\lesssim 1$ to be a typical range of the normalized heat flux values in the solar wind, although that is realistic only at $\beta_{c}\lesssim 1$, because at $\beta_{c}\gtrsim 1$ the normalized heat flux is below a threshold, $q_{e}/q_0\lesssim 1/\beta_{c}$ (see, e.g., Refs. \blue{\cite{Gary1999b,Tong19b:apj,Wilson19b}}).








\begin{table}[]
    \caption{The stability analysis in Sections \ref{sec2} and \ref{sec3} is performed for core and halo parameters indicated in the third column; typical values of these parameters at radial distances of 0.3-4 AU are given in the second column, while several references, where the corresponding spacecraft data analysis was performed, are in the last column. The presented quantities are described in Section \ref{sec2}.}
    \label{tab:first_table}
    \centering
    \begin{tabular}{l@{\hspace*{6mm}}l@{\hspace*{6mm}}ll}
    \toprule
    Quantity & Typical range at 0.3-4 AU & Our Analysis & References\\
    \midrule
     {$n_{c}/n_0$} & {$n_{c}/n_0\gtrsim 0.8$} & {0.95} & McComas et al. \cite{McComas92}\\&&&Maksimovic et al. \cite{Maksimovic05}\\
     & & & Pierrard et al. \cite{Pierrard16}  \\ 
    \midrule
    {$u_{c}/v_A$} & {$|u_{c}|\lesssim 7\;v_{A}$} & {0, 0.25, ..., 10} & Scime et al. \cite{Scime94}\\
    & & & Tong et al. \cite{Tong18}\\ 
    \midrule
    {$\beta_{c}$} & {$\beta_{c}\sim 0.1-10$} & {0.33, 1, 3} & Tong et al. \cite{Tong18,Tong19b:apj}\\
    & & & Wilson et al. \cite{Wilson19b}\\
    & & & \blue{Artemyev et al. \cite{Artemyev18:1au_stats}}\\
    \midrule
    {$T_{h}/T_{c}$} & {$T_{h}/T_{c}\sim 2-10$} & {4, 6, 10} & Maksimovic et al. \cite{Maksimovic05} \\
    & & & Pierrard et al. \cite{Pierrard16}\\
     & & & Wilson et al. \cite{Wilson19b}\\
    \midrule
    {\shortstack{$A_{c}=T_{c\perp}/T_{c||}$ \\ $A_{h}=T_{h\perp}/T_{h||}$}} & {\shortstack{$0.5 \lesssim A_{c}\lesssim \blue{1.2}$ \\ $0.5 \lesssim A_{h}\lesssim 1.5$}} & {\shortstack{$A_{c}=1\;\;\;\;\;\;\;\;\;\;\;\;\;\;\;\;\;\;\;\;\;\;\;\;\;\;$\\ $A_{h}=1.1,.., 1.5$\;\;\;\;\;\;\;\;\;}} & Pierrard et al. \cite{Pierrard16}\\
    & & & \blue{Lazar et al. \cite{Lazar17:AA}}\\
    & & & \blue{Wilson et al. \cite{Wilson19b}}\\
    & & & \blue{Shaaban et al. \cite{Shaaban19:aa}}\\
    \midrule
     {$\kappa$}& {$3 \lesssim \kappa\lesssim 8$} & {3, 5, 7, 9, $\infty$} & Maksimovic et al. \cite{Maksimovic05}\\
     & & & Pierrard et al. \cite{Pierrard16}\\
     & & & Wilson et al. \cite{Wilson19a}\\
    \bottomrule
    \end{tabular}
\end{table}

\section{WHFI and WTAI: effects of the $\kappa-$distribution \label{sec2}}

In this section, we consider effects of \blue{power-law} index of the halo $\kappa-$distribution on growth rates of the whistler heat flux instability (WHFI) and whistler temperature anisotropy instability (WTAI). The effects of the $\kappa-$distribution on growth rates of the WTAI were previously considered in \blue{Refs. \cite{Mace&Sydora10,Lazar13:AA,Lazar19:apss}}, \blue{while until recently} the analysis of effects of the $\kappa-$distribution on the WHFI has been limited to a study of Abraham-Shrauner and Feldman \cite{Abraham-Shrauner77} that was restricted to a single value of the core drift velocity and showed that \blue{Maxwellian halo population} provides the largest growth rates. \blue{In contrast, in the recent analysis of Shaaban {\it et} al.\cite{Shaaban18:mnras} it has been demonstrated that larger growth rates to the WHFI are provided by the halo population described by $\kappa-$distributions, rather than the Maxwell distribution. We demonstrate that conclusions of Refs. \cite{Abraham-Shrauner77} and \cite{Shaaban18:mnras} are valid in some range} of core drift velocity values and depending on parameters of the electron VDF, either the Maxwell or $\kappa-$distributions provide \blue{larger} growth rates to the WHFI. 

\begin{figure*}
    \centering
    \includegraphics[width=0.7\textwidth]{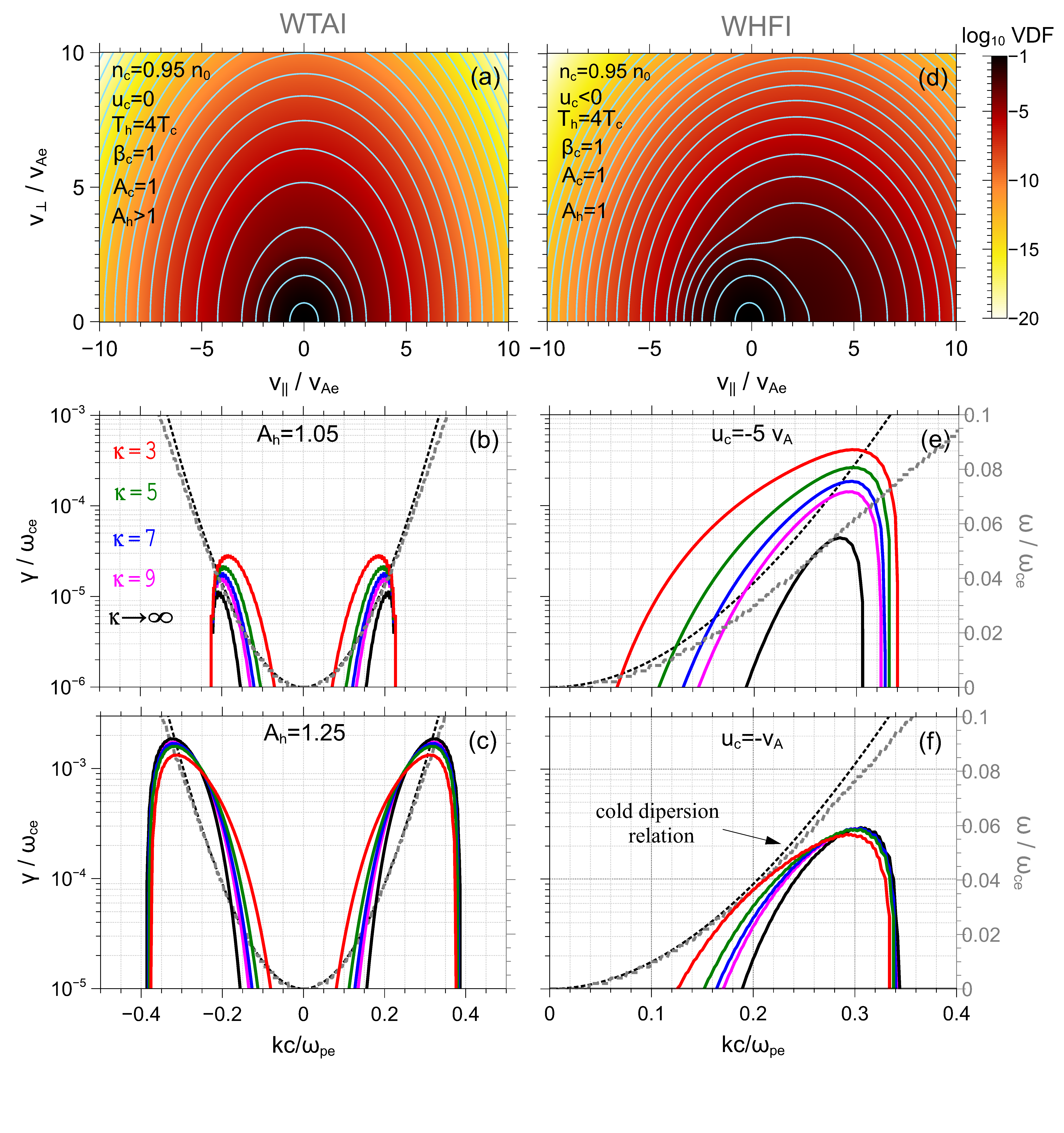}
    \caption{The stability analysis of the whistler temperature anisotropy instability (WTAI) and the whistler heat flux instability (WHFI) of a plasma with cold ions and electron population consisting of a dense thermal core and a tenuous suprathermal halo described by Maxwell and $\kappa-$distributions, respectively. Parameters $n_{\alpha}$, $u_{\alpha}$, $T_{\alpha}$ and $A_{\alpha}$ are density, parallel drift velocity (in the plasma rest frame), parallel temperature and temperature anisotropy (perpendicular to parallel temperature ratio) of the core ($\alpha=c$) and halo ($\alpha=h$) populations; $\beta_{c}=8\pi n_{c}T_{c}/B_0^2$ is the core parallel beta. The upper panels present unstable electron VDFs with corresponding parameters indicated in the panels; velocities parallel and perpendicular to a uniform quasi-static magnetic field, $v_{||}$ and $v_{\perp}$, are normalized to $v_{Ae}=B_0/(4\pi n_0 m_{e})^{1/2}$. The middle and bottom panels present whistler wave growth rates, $\gamma/\omega_{ce}$ versus $kc/\omega_{pe}$, for several values of halo temperature anisotropy $A_{h}$ and core drift velocity $u_{c}/v_{A}$, where $\omega_{ce}$ and $\omega_{pe}$ are electron cyclotron and plasma frequencies, $v_{A}=B_0/(4\pi n_0 m_{i})^{1/2}$ is the Alfv\'{e}n velocity. The growth rates are computed for several values of power-law index $\kappa$ including \blue{$\kappa\rightarrow \infty$}, which corresponds to \blue{Maxwellian halo population}. The panels also \blue{present exact whistler wave dispersion curves} (dashed grey), $\omega/\omega_{ce}$ versus $kc/\omega_{pe}$, \blue{as well as whistler wave dispersion curves (dashed black) valid in a cold plasma at $\omega\gg \omega_{ci}$: $\omega=\omega_{ce} k^2c^2/(k^2c^2+\omega_{pe}^2)$ \cite{Mikhailovskii74,Shklyar04}}.\label{fig1}}
\end{figure*}

Figure \ref{fig1} presents results of the stability analysis of WTAI and WHFI at $\beta_{c}=1$ and $T_{h}=4\;T_{c}$. Panel (a) demonstrates the electron VDF that is unstable to the WTAI due to a perpendicular temperature anisotropy of the halo population. Panels (b) and (c) present the growth rates of the WTAI computed at $A_{h}=1.05$ and $A_{h}=1.25$, and various values \blue{of $\kappa$} including \blue{$\kappa \rightarrow \infty$}, which corresponds to the Maxwell distribution. The Maxwell distribution provides smaller growth rates compared to $\kappa-$distributions at $A_{h}=1.05$, while the opposite can be seen at $A_{h}=1.25$. The different dependence of the growth rate of the WTAI on power-law index $\kappa$ at various electron temperature anisotropies was reported and thoroughly considered in the previous studies \blue{\cite{Mace&Sydora10,Lazar13:AA}}. Noteworthy that the WTAI is capable of producing identical whistler waves propagating parallel and anti-parallel to a local quasi-static magnetic field, because the electron VDF is symmetric with respect to $v_{||}=0$.


Panel (d) demonstrates the electron VDF unstable to the WHFI. The core and halo populations are drifting in the opposite directions, and both populations are isotropic, $A_{c}=1$ and $A_h=1$. Due to the halo drift the contours of the electron VDF at $v_{||}<0$ are in effect similar to those of the electron VDF unstable to the WTAI. Therefore, cyclotron resonant electrons with $v_{||}=(\omega-\omega_{ce})/k$ at $v_{||}<0$ are capable of driving whistler waves propagating parallel to the halo drift and, hence, parallel to the electron heat flux \cite{Gary75,Gary1994a}. In contrast, whistler waves propagating anti-parallel to the electron heat flux are stable. Panels (e) and (f) present the growth rates of the WHFI computed at core drift velocities $u_{c}=-5 v_{A}$ and $u_{c}=-v_{A}$, and various values of \blue{power-law} index $\kappa$. At $u_{c}=-5 v_{A}$ the growth rate is larger for smaller values \blue{of $\kappa$}, so that $\kappa-$distributions provide larger growth rates than the Maxwell distribution. On the other hand, at $u_{c}=-v_{A}$ the largest growth rate is provided by the Maxwell distribution. Thus, similarly to the WTAI, larger growth rate is be provided either by the Maxwell or $\kappa-$distribution \blue{depending on parameters of the electron VDF}. To further quantify effects of power-law index $\kappa$ on the WHFI, we have computed the growth rate $\gamma_{\rm max}$, frequency $\omega_{\rm max}$ and wave number $k_{\rm max}$ of the fastest growing whistler waves at various values of parameters $\beta_{c}$ and $T_{h}/T_{c}$.

\begin{figure*}
    \centering
    \includegraphics[width=1.0\textwidth]{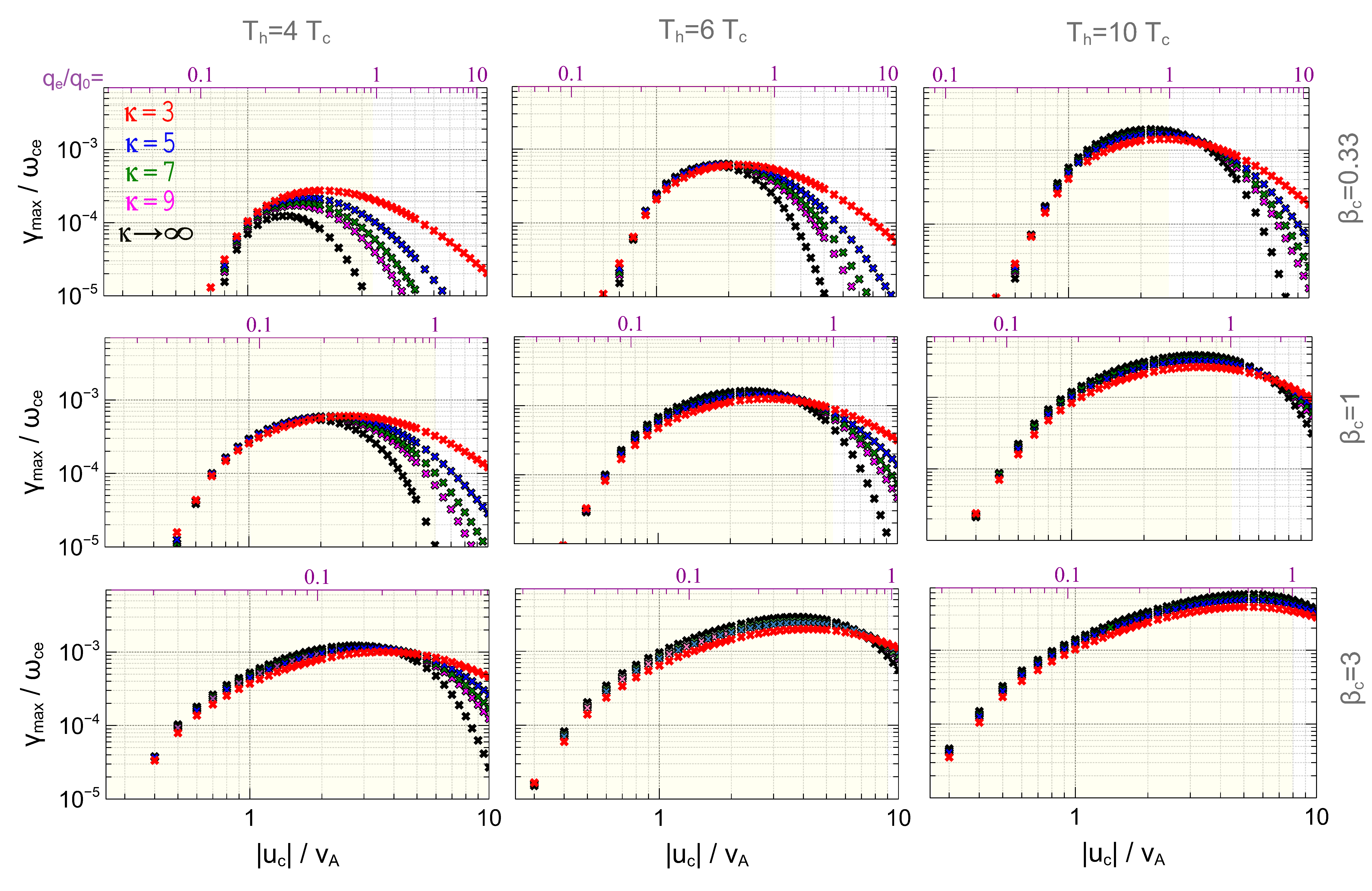}
    \caption{The maximum growth rate $\gamma_{\rm max}/\omega_{ce}$ of the WHFI in dependence on core drift velocity $u_{c}/v_{A}$ and electron heat flux $q_{e}/q_0$ (upper horizontal axes), where $v_{A}$ is the Alv\'{e}n velocity and $q_0$ is \blue{the free-streaming heat flux value\cite{Cowie77,Gary1999b} defined as $q_{0}=1.5 \;n_0\;T_{e}\;(2T_{e}/m_{e})^{1/2}$, where $n_0=n_c+n_h$ is the total electron density, $T_{e}=(n_cT_c+n_hT_h)/n_0$}. The growth rates were computed for various values of power-law index $\kappa$ of the halo $\kappa-$distribution (indicated by color). The various panels corresponds to maximum growth rates computed at various $(\beta_{c},T_{h}/T_{c})$. In all computations the core and halo populations were isotropic ($A_{c}=1$ and $A_{h}=1$) and the density of the core population is $n_{c}=0.95\;n_0$. The shaded regions indicate the range of electron heat flux values typical of the solar wind, i.e. $q_{e}\lesssim q_0$ \blue{\cite{Tong19b:apj,Wilson19b}}.\label{fig2}}
\end{figure*}

\begin{figure*}
    \centering
    \includegraphics[width=1.0\textwidth]{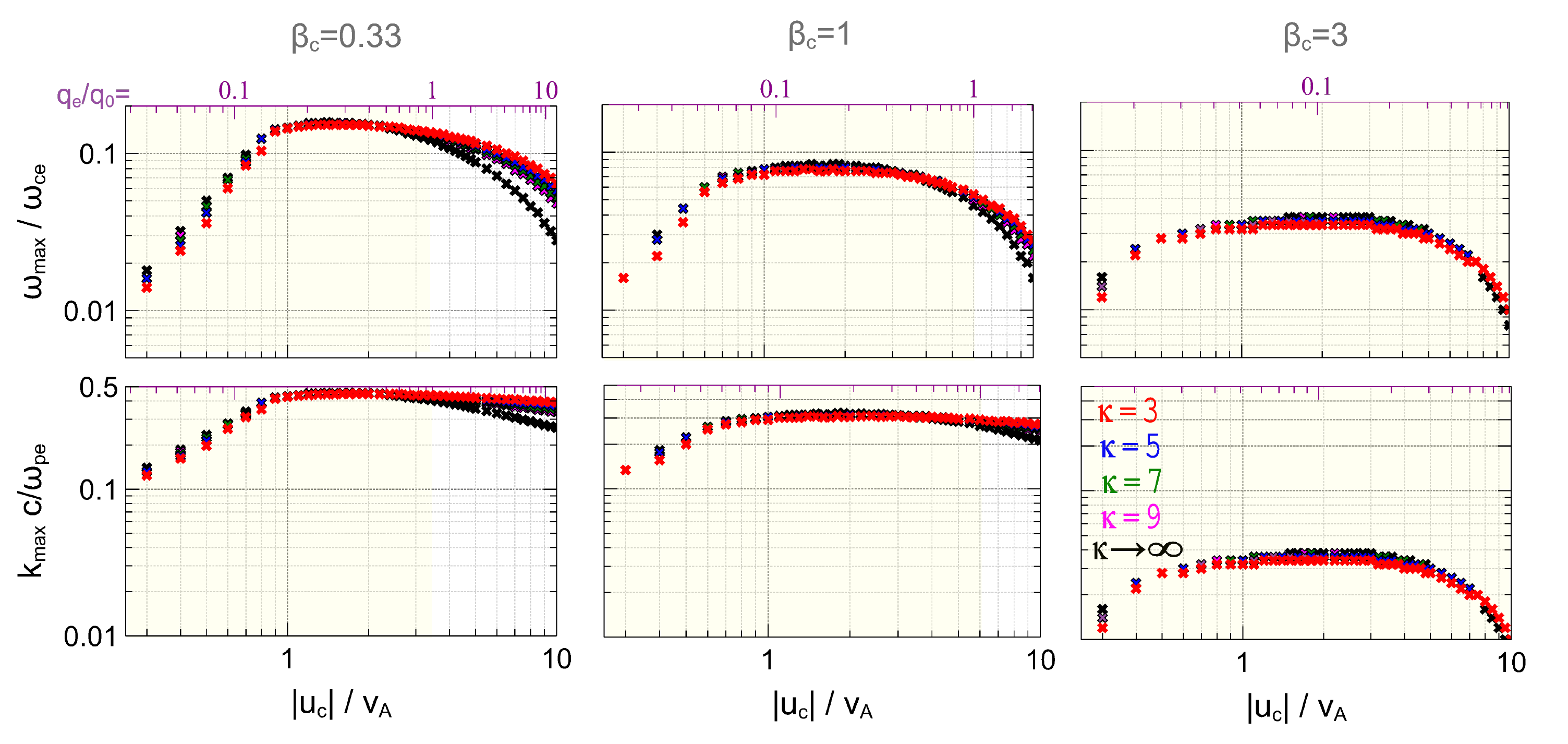}
    \caption{The frequency $\omega_{\rm max}/\omega_{ce}$ and wave number $k_{\rm max}c/\omega_{pe}$ of the fastest growing whistler waves in dependence on $u_{c}/v_{A}$ and $q_{e}/q_0$ (upper horizontal axes) at $T_{h}/T_{c}=4$ and various values of \blue{core electron beta $\beta_{c}$} and \blue{power-law index} $\kappa$. The corresponding growth rates are shown in the first column of Figure \ref{fig2}. Only panels corresponding to $T_{h}/T_{c}=4$ are demonstrated, because that parameter does not critically affect the frequency and wave number of the fastest growing whistler waves (see Section \ref{sec3} for details). The shaded regions indicate the range of electron heat flux values typical of the solar wind, i.e. $q_{e}\lesssim q_0$ \blue{\cite{Tong19b:apj,Wilson19b}}. \label{fig3}}
\end{figure*}

Figure \ref{fig2} presents the maximum growth rate $\gamma_{\rm max}/\omega_{ce}$ of the WHFI in dependence on $u_{c}/v_{A}$ and $q_{e}/q_0$ computed at various values of parameters $\beta_{c}$, $T_{h}/T_{c}$ and $\kappa$ (Table \ref{tab:first_table}). Both core and halo populations were assumed to be isotropic, $A_{c}=1$ and $A_{h}=1$. The range of electron heat flux values typical of the solar wind, $q_{e}/q_0\lesssim 1$, is indicated by shaded regions in each of the panels. First, at a given $(\kappa,\beta_{c},T_{h}/T_{c})$ the maximum growth rate of the WHFI is a non-monotonous function of $u_{c}/v_{A}$ and $q_{e}/q_0$ in accordance with the original analysis of Gary et al. \cite{Gary85} that was restricted to the Maxwell distribution (\blue{$\kappa\rightarrow\infty$}). Second, in accordance with Gary et al. \cite{Gary85}, larger values of $\beta_{c}$ and $T_{h}/T_{c}$ result in larger growth rates. The novel feature of the WHFI demonstrated by Figure \ref{fig2} is that at a given value of $u_{c}/v_{A}$ and $q_{e}/q_0$ the maximum growth rate can either increase or decrease with the increase of the power-law index $\kappa$. The Maxwell distribution provides the largest growth rates at \blue{core drift velocities $|u_{c}|/v_{A}$ below a threshold value that monotonously depends on $\beta_{c}$ and $T_{h}/T_{c}$, while at core drift velocities above that threshold value} the largest growth rates are \blue{provided} by the $\kappa-$distribution with the lowest power law index. \blue{For instance, at $\beta_{c}=1$ and $T_{h}/T_{c}=6$ the threshold value is $|u_{c}|\approx 3\;v_{A}$.} Noteworthy that \blue{at core drift velocities below the threshold} value the growth rates provided by the Maxwell distribution exceed those provided by $\kappa-$distributions by less than a factor of two. On the other hand, at \blue{core drift velocities above the threshold value} the $\kappa-$distributions provide growth rates more than an order of magnitude larger than the Maxwell distribution.

Figure \ref{fig3} presents the frequency $\omega_{\rm max}/\omega_{ce}$ and wave number $k_{\rm max}c/\omega_{pe}$ of the fastest growing whistler waves at various $\beta_{c}$ and $\kappa$ values. The specific value of $T_{h}/T_{c}$ does not critically affect the frequency and wave number of the fastest growing whistler waves (see Section \ref{sec3}) that is why in Figure \ref{fig3} we have limited the presentation to $T_{h}/T_{c}=4$. First, the frequency and wave number of the fastest growing whistler waves are non-monotonously dependent on $u_{c}/v_{A}$. Second, in accordance with Gary et al. \cite{Gary85} parameter $\beta_{c}$ strongly affects frequency and wave numbers of the fastest growing whistler waves. The novel feature of the WHFI demonstrated by Figure \ref{fig3} is that the specific value of the power-law index $\kappa$ does not significantly affect the frequency and wave number of the fastest growing whistler waves. 

A brief summary of this section is that \blue{the maximum growth rate of the WHFI is not a monotonous function of power-law index $\kappa$} and that power-law index $\kappa$ can critically affect the growth rates of the WHFI \blue{(the latter is in accordance with recent analysis of Ref. \cite{Shaaban18:mnras})}, but does not essentially affect the frequency and wave number of the fastest growing whistler waves. Because larger growth rates \blue{result in} larger saturated amplitudes of whistler waves according to the quasi-linear scaling relation (see Section \ref{sec3}), power law index $\kappa$ is expected to affect \blue{saturated} amplitudes of whistler waves, especially \blue{at core drift velocity values above some threshold dependent on $\beta_{c}$ and $T_{h}/T_{c}$.}

\section{Parallel and anti-parallel whistler waves\label{sec3}}

\subsection{Linear growth rates}

\begin{figure*}
    \centering
    \includegraphics[width=0.6\textwidth]{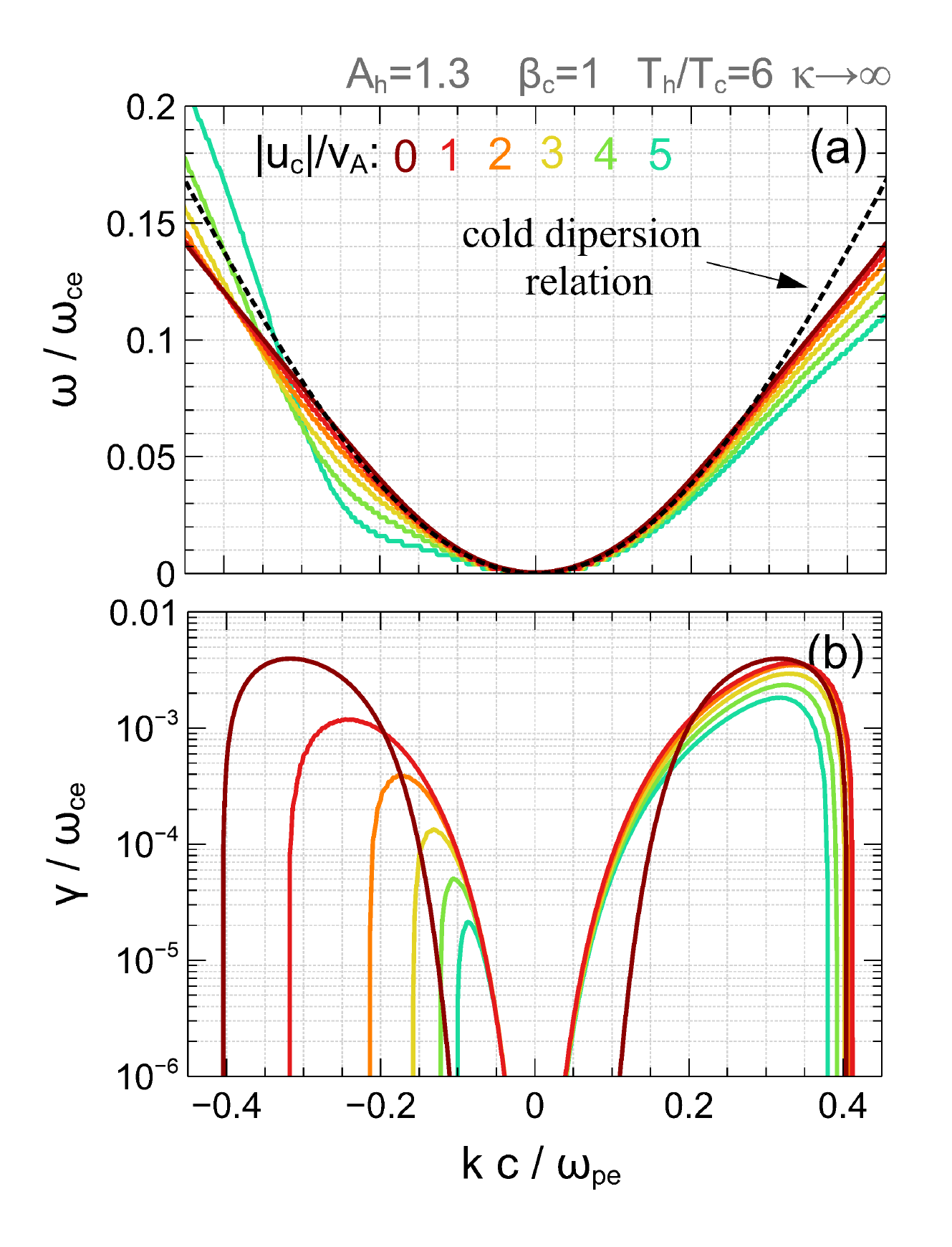}
    \caption{The stability analysis of parallel and anti-parallel whistler waves at a fixed value of the halo temperature anisotropy, $A_{h}=1.3$, and various values of core drift velocity $|u_{c}|/v_{A}$: (a) dispersion curves of parallel ($kc/\omega_{pe}>0$) and anti-parallel ($kc/\omega_{pe}<0$) whistler waves; dashed black curves represent the whistler wave dispersion curves in a cold plasma \blue{at $\omega\gg \omega_{ci}$} \blue{\cite{Mikhailovskii74,Shklyar04}}: $\omega=\omega_{ce} k^2c^2/(k^2c^2+\omega_{pe}^2)$; (b) the growth rates of parallel and anti-parallel whistler waves; the growth rate computed at $u_{c}=0$ corresponds to the whistler temperature anisotropy instability (WTAI), which produces identical parallel and anti-parallel whistler waves. The stability analysis was performed at $\beta_{c}=1$, $T_{h}/T_{c}=6$ and \blue{$\kappa\rightarrow\infty$}.\label{fig4}}
\end{figure*}

\begin{figure*}
    \centering
    \includegraphics[width=0.65\textwidth]{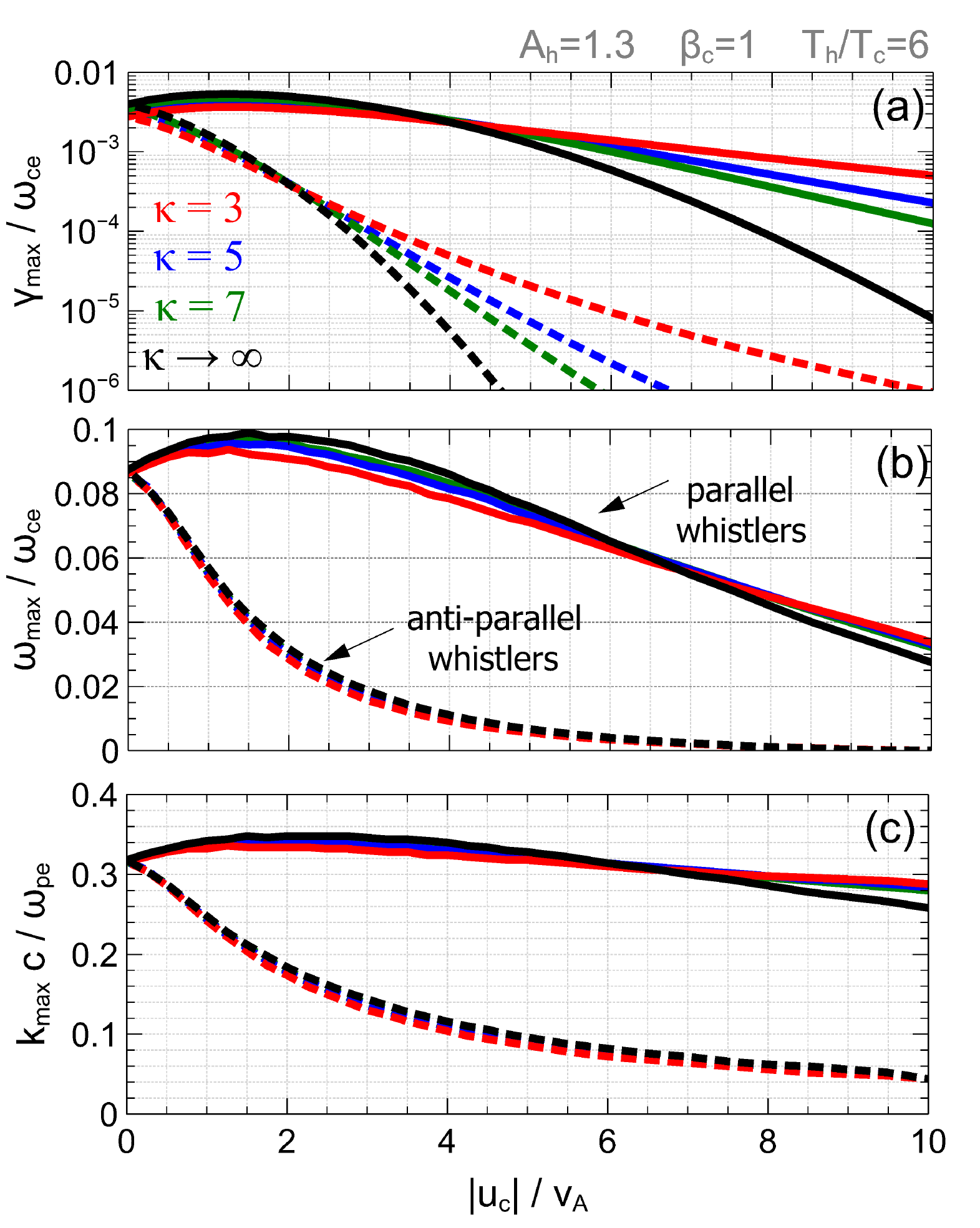}
    \caption{The \blue{properties} of the fastest growing parallel and anti-parallel whistler waves computed at a fixed value of the halo temperature anisotropy, $A_{h}=1.3$, and various values of power-law index $\kappa$ and core drift velocity $|u_{c}|/v_{A}$: (a) the growth rate $\gamma_{\rm max}/\omega_{ce}$, (b) frequency $\omega_{\rm max}/\omega_{ce}$ and (c) wave number $|k_{\rm max}|c/\omega_{pe}$. The parameters of parallel and anti-parallel whistler waves are shown by solid and dashed curves, respectively. The other parameters are $\beta_{c}=1$ and $T_{h}/T_{c}=6$.\label{fig5}}
\end{figure*}

\begin{figure*}
    \centering
    \includegraphics[width=0.65\textwidth]{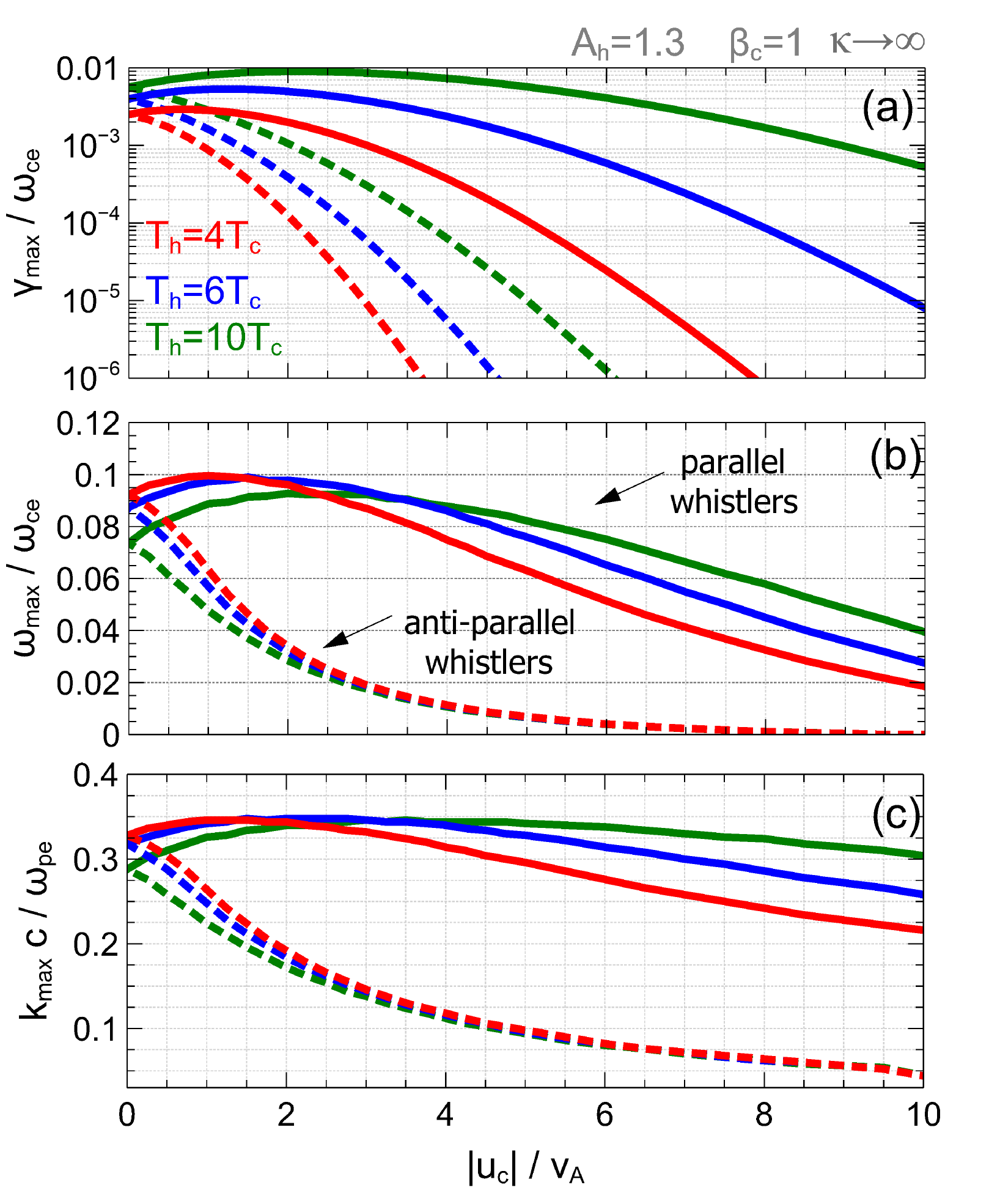}
    \caption{The properties of the fastest growing parallel and anti-parallel whistler waves computed at a fixed \blue{value} of the halo temperature anisotropy, $A_{h}=1.3$, \blue{and various values of core drift velocity $|u_{c}|/v_{A}$ and halo to core parallel temperature ratio $T_{h}/T_{c}$.} The format of the figure is identical to that of Figure \ref{fig5}. The other parameters are $\beta_{c}=1$ and \blue{$\kappa\rightarrow \infty$}.\label{fig6}}
\end{figure*}

\begin{figure*}
    \centering
    \includegraphics[width=0.65\textwidth]{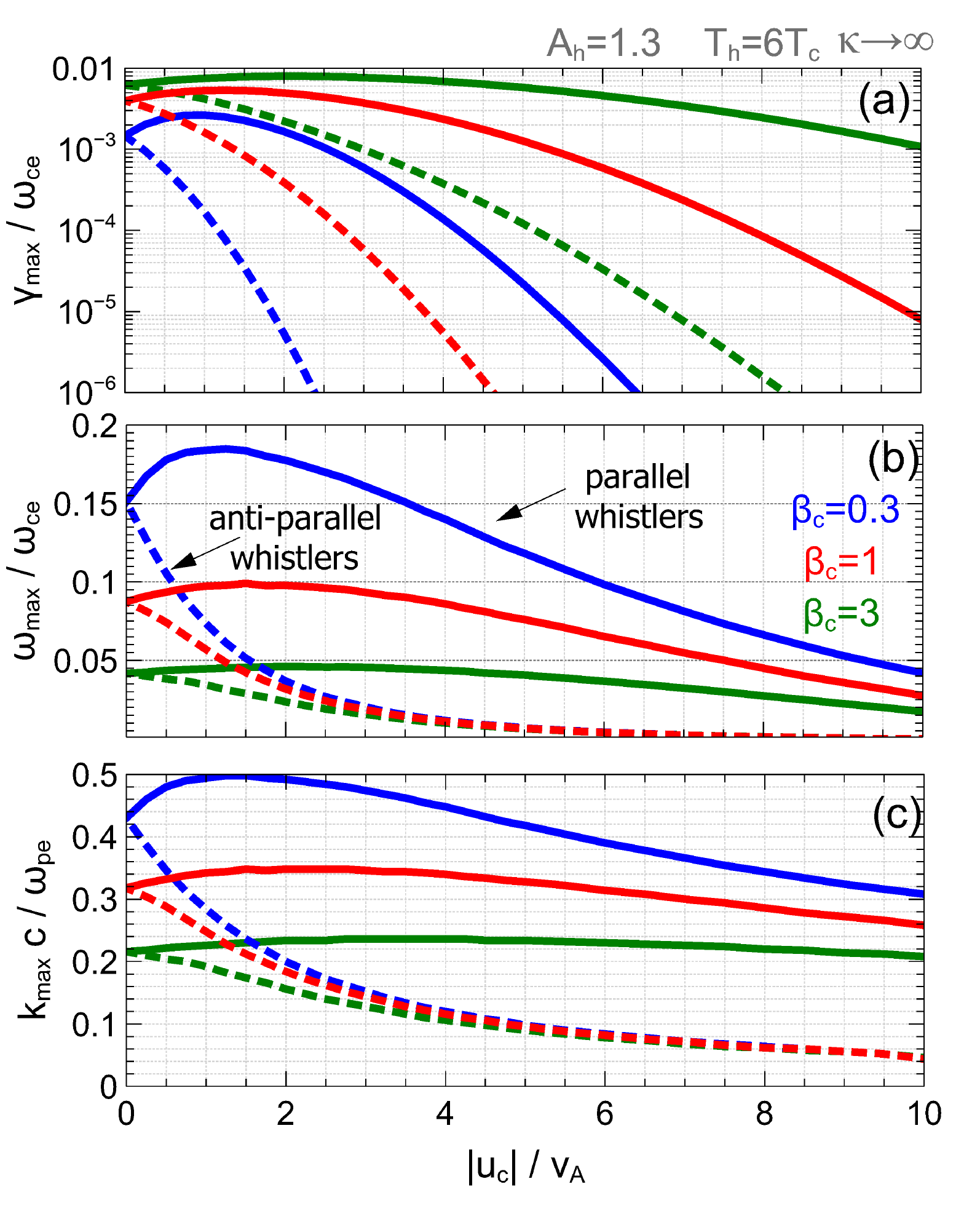}
    \caption{The properties of the fastest growing parallel and anti-parallel whistler waves computed at a fixed value of the halo temperature anisotropy, $A_{h}=1.3$, and various values of core parallel beta parameter $\beta_{c}$ and core drift velocity $|u_{c}|/v_{A}$. The format of the figure is identical to that of Figure \ref{fig5}. The other parameters are $T_{h}/T_{c}=6$ and \blue{$\kappa\rightarrow \infty$}.\label{fig7}}
\end{figure*}

\begin{figure*}
    \centering
    \includegraphics[width=1.0\textwidth]{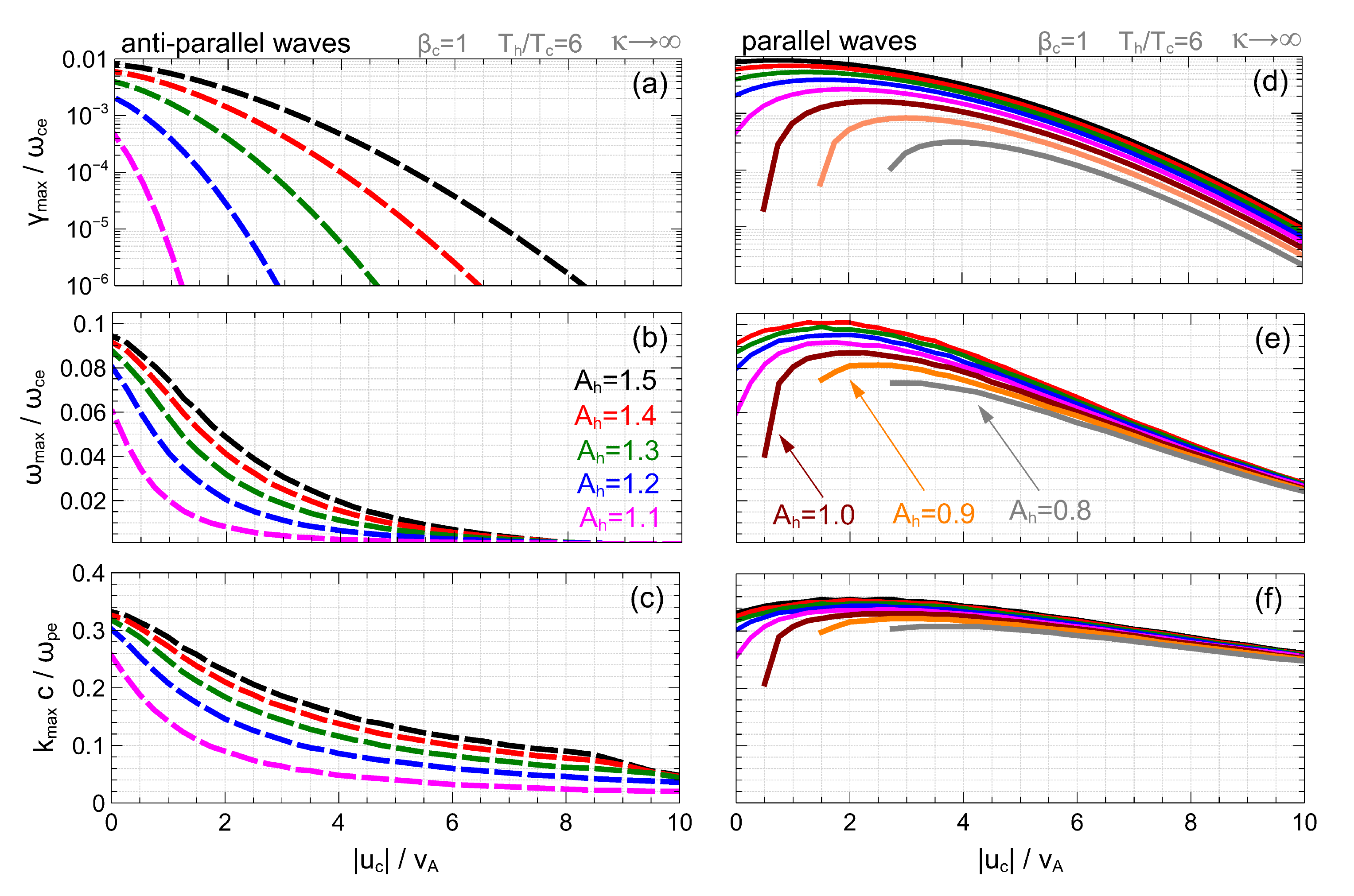}
    \caption{\blue{The properties of the fastest growing parallel and anti-parallel whistler waves computed at various values of the halo temperature anisotropy $A_{h}=T_{h\perp}/T_{h||}$ and core drift velocity $|u_{c}|/v_{A}$. The properties of parallel whistler waves in panels (e)-(d) are demonstrated for $0.8 \leq A_{h}\leq 1.5$, while the properties of anti-parallel whistler waves are shown for $1.1\leq A_{h}\leq 1.5$, because anti-parallel whistler waves are stable at $A_h\leq 1$. The format of panels (a)-(c) and (e)-(d) is identical} to that of Figure \ref{fig5}. The other parameters are $\beta_{c}=1$, $T_{h}/T_{c}=6$ and \blue{$\kappa\rightarrow\infty$}.\label{fig8}}
\end{figure*}

In the presence of isotropic or parallel-anisotropic halo population the WHFI is capable of producing only whistler waves propagating parallel to the electron heat flux \cite{Gary75,Gary1994a}. Whistler waves propagating parallel and anti-parallel to the electron heat flux can be unstable in the presence of a perpendicular temperature anisotropy of the halo population, $A_{h}=T_{h\perp}/T_{h||}>1$. In this section, we present a quantitative analysis of effects of parameters $\beta_{c}$, $T_{h}/T_{c}$ and $\kappa$, \blue{whose} typical values are indicated in Table \ref{tab:first_table}, on the growth rate, frequency and wave number of parallel and anti-parallel whistler waves unstable at \blue{various values of} core drift velocity $|u_{c}|/v_{A}$ and perpendicular halo temperature anisotropy $A_{h}$. In each of Figures \ref{fig4}-\ref{fig8} we demonstrate effects of a particular parameter on \blue{properties of unstable whistler waves}, while other parameters have default values, which are $A_{h}=1.3$, $\beta_{c}=1$, $T_{h}/T_{c}=6$ and \blue{$\kappa\rightarrow \infty$}. \blue{In Figure \ref{fig8} we demonstrate effects of the halo temperature anisotropy on unstable anti-parallel and parallel whistler waves for $1\leq A_{h}\leq 1.5$ and $0.8\leq A_{h}\leq 1.5$, respectively.}.

Figure \ref{fig4} presents the stability analysis of parallel and anti-parallel whistler waves performed at a fixed value of the halo temperature anisotropy, $A_{h}=1.3$, and various values of the core drift velocity $u_{c}/v_{A}$. The other parameters are $\beta_{c}=1$, $T_{h}/T_{c}=6$ and \blue{$\kappa\rightarrow \infty$}. Panel (a) shows that the presence of \blue{core and, hence, halo drifts} affects whistler wave dispersion curves. Panel (b) presents whistler wave growth rates. \blue{At negligible values of core and halo drifts}, identical parallel and anti-parallel whistler waves are unstable to WTAI. The presence of \blue{core and halo} drifts breaks that symmetry between parallel and anti-parallel propagation resulting in smaller maximum growth rates of anti-parallel whistler waves. For instance, at $|u_{c}|=3v_{A}$ the fastest growing parallel whistler waves have frequency $\omega_{\rm max}/\omega_{ce}\approx 0.07$, wave number $k_{\rm max}c/\omega_{pe}\approx 0.35$ and growth rate $\gamma_{\rm max}/\omega_{ce}\approx 3\cdot 10^{-3}$, while the fastest growing anti-parallel whistler wave is at lower frequency, $\omega_{\rm max}/\omega_{ce}\approx 0.02$, \blue{smaller} wave number, $k_{\rm max}c/\omega_{pe}\approx -0.15$, and has \blue{one} order of magnitude smaller growth rate, $\gamma_{\rm max}/\omega_{ce}\approx 10^{-4}$. In what follows, we compare properties $\gamma_{\rm max}/\omega_{ce}$, $\omega_{\rm max}/\omega_{ce}$ and $|k_{\rm max}|c/\omega_{pe}$ of the fastest growing parallel and anti-parallel whistler waves at various \blue{values of parameters} $\beta_{c}$, $T_{h}/T_{c}$ and $\kappa$ .

Figure \ref{fig5} presents comparison \blue{of the} fastest growing parallel and anti-parallel whistler waves at a fixed value of the halo temperature anisotropy, $A_{h}=1.3$, and various values of power-law index $\kappa$ and core drift velocity $u_{c}/v_{A}$. The other parameters are $\beta_{c}=1$ and $T_{h}/T_{c}=6$. Panel (a) shows that depending on the core drift velocity the growth rates of both parallel and anti-parallel whistler waves can \blue{increase or decrease with increasing value of power-law index} $\kappa$. Similarly to the WHFI considered in Section \ref{sec2}, \blue{the $\kappa-$distribution with the lowest power-law index provides the largest growth rate at core drift velocities larger than some threshold value, while the Maxwell distribution provides the largest growth rates at core drift velocities below that threshold value. At $\beta_{c}=1$ and $T_{h}/T_{c}=6$ assumed in Figure \ref{fig5} these threshold values are $|u_{c}|\approx 2\;v_{A}$ for anti-parallel whistler waves and $|u_{c}|\approx 4\;v_{A}$ for parallel whistler waves. We note that similarly to analysis of the WHFI in Section \ref{sec2} these threshold values are generally dependent on $\beta_{c}$ and $T_{h}/T_{c}$.} Panels (b) and (c) show that \blue{power-law index} $\kappa$ does not \blue{significantly} affect the frequency and wave number of the fastest growing whistler waves. Panels (a)-(c) also demonstrate that the properties of the fastest growing parallel and anti-parallel whistler waves are different. There is a factor of a few difference between the growth rates of parallel and anti-parallel whistler waves \blue{at low core drift velocities}, but already at $|u_{c}|\gtrsim 2\;v_{A}$ the growth rates of anti-parallel whistler waves are \blue{more than one order} of magnitude smaller than those of the parallel whistler waves. \blue{At $|u_{c}|\gtrsim 2\;v_{A}$,} the frequencies and wave numbers of the anti-parallel whistler waves are a few times \blue{smaller} than those of the parallel whistler waves.

Figure \ref{fig6} presents comparison of the fastest growing parallel and anti-parallel whistler waves at a fixed value of the halo temperature anisotropy, $A_{h}=1.3$, \blue{and various values of core drift velocity $u_{c}/v_{A}$ and halo to core parallel temperature ratio $T_{h}/T_{c}$.} The other parameters are $\beta_{c}=1$ and \blue{$\kappa\rightarrow \infty$}. Panel (a) shows that larger values of $T_{h}/T_{c}$ result in larger growth rates of both parallel and anti-parallel whistler waves. Parameter $T_{h}/T_{c}$ rather critically affects the growth rates, namely already at $|u_{c}|\gtrsim 2\;v_{A}$, the growth rates corresponding to $T_{h}/T_{c}=4$ and 10 differ by more than \blue{one} order of magnitude. Panels (b) and (c) demonstrate that parameter $T_{h}/T_{c}$ does not critically affect frequencies and wave numbers of both parallel and anti-parallel whistler waves.

Figure \ref{fig7} presents comparison of the properties of the fastest growing parallel and anti-parallel whistler waves at a fixed value of the halo temperature anisotropy, $A_{h}=1.3$, and various values of \blue{core parallel} beta parameter $\beta_{c}$ and core drift velocity $u_{c}/v_{A}$. The other parameters are $T_{h}/T_{c}=6$ and \blue{$\kappa\rightarrow \infty$}. Panel (a) shows that, similarly to the WHFI considered in Section \ref{sec2}, the maximum growth rates of parallel and anti-parallel whistler waves \blue{are} critically dependent on $\beta_{c}$ and larger values of $\beta_{c}$ result in larger growth rates. Panels (b) and (c) show that, similarly to the WHFI, the frequency and wave number of both parallel and anti-parallel whistler waves are dependent on $\beta_{c}$. Generally, larger values of $\beta_{c}$ result in the fastest growing whistler waves at lower frequencies and \blue{smaller} wave numbers. In the range of $\beta_{c}=0.33-3$ the frequency and wave numbers of the fastest growing whistler waves vary by a factor of a few.

\blue{Figure \ref{fig8} presents the properties of the fastest growing whistler waves at various values of halo temperature anisotropy $A_{h}$ and core drift velocity $u_{c}/v_{A}$. The properties of anti-parallel whistler waves in panels (a)-(c) are shown for $1\leq A_{h}\leq 1.5$, because anti-parallel whistler waves are stable at $A_{h}\leq 1$, while the properties of parallel whistler waves in panels (e)-(d) are demonstrated for $0.8\leq A_{h}\leq 1.5$. The other parameters are $\beta_{c}=1$, $T_{h}/T_{c}=6$ and $\kappa\rightarrow \infty$. Panel (a) shows that, as expected, the growth rates of anti-parallel whistler waves increase with increasing halo temperature anisotropy, while at a fixed halo temperature anisotropy the growth rate decreases with increasing core drift velocity $|u_{c}|/v_{A}$. Panels (b) and (c) show that larger values of the halo temperature anisotropy result in larger frequencies and wave numbers of unstable anti-parallel whistler waves. The properties of unstable parallel whistler waves in panels (e)-(d) vary with increasing halo temperature anisotropy $A_{h}$ in similar way, except that at a fixed halo temperature anisotropy whistler wave properties are generally dependent on $|u_{c}|/v_{A}$ non-monotonously. The increase of the growth rates, frequencies and wave numbers for increasing halo temperature anisotropy is in accordance with similar characteristics of WTAI \cite{Kennel66,Mace&Sydora10,Lazar13:AA}. The critical feature is that due to a finite electron heat flux parallel whistler waves can be unstable even in the presence of parallel-anisotropic halo electrons, $A_{h}\leq 1$. The comparison of growth rates computed for $A_{h}=0.8$ and $0.9$ in panel (e) shows that larger core drift velocities $|u_{c}|/v_{A}$ and, hence, heat flux values $q_{e}/q_0$ are necessary to have unstable parallel whistler waves at larger parallel temperature anisotropies. That is in accordance with the early stability analysis of the WHFI by \cite{Gary75} and \cite{Abraham-Shrauner77}, who showed that $T_{h||}/T_{h\perp}>1$ may quench the WHFI, and consistent with recent spacecraft measurements by \cite{Tong2019a}, who confirmed this stabilization effect experimentally. The stabilization effect of the halo parallel temperature anisotropy can explain the intermittent appearance of whistler waves in the solar wind \cite{Tong2019a,Tong19b:apj}. Finally, the comparison of panels (b) and (c) to panels (f) and (d) shows that for a wide range of typical halo temperature anisotropies parallel whistler waves have higher frequencies and larger wave numbers than anti-parallel whistler waves.}



\subsection{Saturated amplitudes of whistler waves}

\blue{There have been recently reported PIC simulations of nonlinear evolution of parallel whistler waves produced by the WHFI \cite{Kuichev19,Lopez19} and quasi-linear simulations of nonlinear evolution of parallel whistler waves unstable in the presence of electron heat flux and core and halo temperature anisotropies \cite{Lazar18:jgr,Lazar19:apss,Shaaban19aa,Shaaban19:mnras,Shaaban20:mnras}. These simulations have provided valuable information on evolution of the electron velocity distribution function due to scattering of electrons by unstable parallel whistler waves and demonstrated that for electron parameters typical of the pristine solar wind parallel whistler waves should saturate at rather low amplitudes, $B_{w}/B_0\sim 0.01$. These amplitudes are consistent with spacecraft measurements by Tong et al. \cite{Tong19b:apj}, who showed that statistically whistler wave amplitudes are $B_{w}/B_0\lesssim 0.01$. We stress that these are amplitudes of whistler waves in the pristine solar wind, while whistler waves with larger amplitudes can be expected in a disturbed solar wind, i.e., for example, around interplanetary shock waves \cite{Kennell82,Coroniti82,Wilson13,Liu18:apj}. Although quasi-linear computations including effects of anti-parallel whistler waves would be desirable to perform in the future, they are actually not necessary to estimate saturated amplitudes of anti-parallel whistler waves by the order of magnitude.}

\blue{The quasi-linear analysis of WTAI by Tao et al. \cite{Tao17} and PIC simulations of the WHFI by Kuzichev et al.\cite{Kuichev19} showed that there is a remarkable scaling relation, which shows that the saturated amplitude of unstable parallel whistler waves depends on the initial maximum increment $\gamma_{\rm max}$ of whistler waves
\begin{eqnarray}
B_w/B_0\propto (\gamma_{\rm max}/\omega_{ce})^{\alpha},  
\label{eq:qlt}
\end{eqnarray}
where $\alpha$ is about 0.7. Although Kuzichev et al. \cite{Kuichev19} inferred that scaling relation using PIC simulations, the nonlinear evolution of the WHFI in those simulations is actually quasi-linear and, hence, Eq. (\ref{eq:qlt}) can be referred by the quasi-linear scaling relation. The quasi-linear scaling relation is expected to be applicable in the solar wind, because spacecraft measurements and estimates by \cite{Tong19b:apj} demonstrated that the quasi-linear theory is indeed applicable to whistler waves in the solar wind. Moreover, using the quasi-linear scaling relation Kuzichev et al. \cite{Kuichev19} conclusively explained the dependence of $B_{w}/B_0$ on $q_{e}/q_0$ and $\beta_{e}$ revealed in the spacecraft measurements by \cite{Tong19b:apj}.} 

\blue{The quasi-linear scaling relation allows estimating saturated amplitudes of anti-parallel whistler waves expected in the solar wind. Anti-parallel whistler waves typically have growth rates smaller than parallel whistler waves and, particularly, at $|u_{c}|\gtrsim 2\;v_{A}$, anti-parallel whistler waves have growth rates of one order of magnitude smaller than parallel whistler waves (see Figures \ref{fig4}-\ref{fig7}). Therefore, the quasi-linear scaling relation indicates that anti-parallel whistler waves should saturate at amplitudes of about one order of magnitude smaller than parallel whistler waves. Taking into account that parallel whistler waves saturate at amplitudes $B_{w}\lesssim 0.01\;B_0$, we conclude that anti-parallel whistler wave in the pristine solar wind are expected to saturate at amplitudes $B_{w}\lesssim 10^{-3}\;B_0$.}

\section{Discussion and conclusion\label{sec4}}

\blue{In this paper, we have estimated and compared properties of parallel and anti-parallel whistler waves that can be unstable in the solar wind due to the presence of the electron heat flux and halo temperature anisotropy.} First, the presented analysis is relevant to a slow solar wind not very close to the Sun, because in that case the electron VDF can be adequately fitted by a combination of core and halo populations \cite{Feldman75,Maksimovic97,Pierrard16,Tong2019a}. \blue{In a fast solar wind around 1 AU or close to the Sun \cite{Rosenbauer77,Pilipp87,Stverak09,Horaites18:mnras,Halekas19}, there is an additional beam-like population (strahl) propagating typically anti-sunward, which contribution to instability of anti-parallel whistler waves could not be neglected. Second, the presented analysis is focused on whistler waves with frequencies much higher than ion cyclotron frequency, so that the contribution of resonant ions could be neglected. Therefore, the presented stability analysis is not relevant to low-frequency fast-magnetosonic/whistler waves recently measured aboard Parker Solar Probe, which are driven by ion beams or ion parallel temperature anisotropy \cite{Bale19:nature,Bowen20:apjs,Verniero20}.}

We have addressed effects of power-law index $\kappa$ of the halo $\kappa-$distribution on the growth rates of parallel whistler waves driven by the WHFI. Along with known properties of the WHFI, i.e. effects of $\beta_{c}$ and $T_{h}/T_{c}$ on growth rates and other properties of unstable whistler waves \cite{Gary85}, we have demonstrated that power-law index $\kappa$ does not critically affect the frequency and wave number of the fastest growing whistler waves, but can critically affect the growth rates \blue{(in accordance with analysis of Shaaban {\it et} al.\cite{Shaaban18:mnras})}. \blue{The growth rate is not monotonously dependent on power-law index $\kappa$}, i.e. depending on \blue{the} core drift velocity, the growth rate can either increase or decrease with increasing $\kappa$ value. In principle, $\kappa$ values typical of the solar wind provide a few times smaller growth rates than the Maxwell distribution at core drift velocities below \blue{a threshold value that depends on $\beta_{c}$ and $T_{h}/T_{c}$}, but more than one order of magnitude larger growth rates at core drift velocities larger than \blue{that threshold value (Figure \ref{fig2})}.


We have estimated and compared properties of the fastest growing parallel and anti-parallel whistler waves unstable in the presence of halo temperature anisotropies and core drift velocities typical of the pristine solar wind (Table \ref{tab:first_table}). We have demonstrated that the growth rates of parallel and anti-parallel whistler waves are larger for larger values of $\beta_{c}$ and $T_{h}/T_{c}$ and critically dependent on the power-law index $\kappa$ in a fashion similar to that of the WHFI. The frequency and wave number of the parallel and anti-parallel whistler waves are most strongly dependent on $\beta_{c}$ and temperature anisotropy $A_{h}$, while essentially independent of $\kappa$ and $T_{h}/T_{c}$. Anti-parallel whistler waves have growth rates smaller than \blue{those of} parallel whistler waves \blue{and, specifically, smaller by more than one} order of magnitude already at $|u_{c}|\gtrsim 2\;v_{A}$. In addition, anti-parallel whistler waves are expected to have a few times lower frequencies and smaller wavenumbers compared to parallel whistler waves. \blue{We have used the quasi-linear scaling relation reported by tao et al. \cite{Tao17} and Kuzichev et al. \cite{Kuichev19} to argue that anti-parallel whistler waves saturate at amplitudes of about one order of magnitude smaller than parallel whistler waves. Therefore, anti-parallel whistler waves in the pristine solar wind are expected to have amplitudes $B_{w}\lesssim 10^{-3}\;B_0$.}

\blue{The presented analysis of anti-parallel whistler waves has been stimulated by the recent simulations, which showed that parallel whistler waves produced by the WHFI cannot regulate the electron heat conduction in the solar wind \cite{Roberg-Clark:2016,Kuichev19,Lopez19,Shaaban19:mnras}. Whether anti-parallel whistler waves can regulate the electron heat conduction in the solar wind is an open question, but a few comments are in order. First, although anti-parallel whistler waves are expected to saturate at relatively low amplitudes, this should not be considered as an indication that these whistler waves are incapable of affecting the electron heat flux in the solar wind. The saturated amplitudes determine the time scale of the saturation of the instability, but not the effects of whistler waves on the electron velocity distribution function. Second, quasi-linear computations of Vocks et al. \cite{Vocks05} showed that anti-parallel whistler waves can efficiently scatter suprathermal electrons in the solar wind and, hence, may be efficient in regulating the electron heat flux (see also Ref. \cite{Vocks12}). Third, in a fast solar wind or close to the Sun the physics of the heat flux regulation can be dominated by oblique whistler waves driven by the strahl population \cite{komarov_2018,Vasko2019a,Verscharen19:apj,Roberg-Clark19} (see Refs. \cite{Breneman10,Cattell20:arxiv} for observations of oblique whistler waves), but in a slow solar wind at sufficiently large distances from the Sun whistler waves propagating anti-parallel to the electron heat flux is a rather plausible wave activity to be involved into the electron heat flux regulation process. The quantitative analysis of the effects of anti-parallel whistler waves on the electron heat flux requires a separate study.}

There are several valuable implications of the results presented in this paper. First, the presented stability analysis demonstrates the range of core and halo parameters, which provides sufficiently large \blue{increments of anti-parallel whistler waves} to be used in numerical PIC simulations. For instance, the presented analysis shows that \blue{at core drift velocities larger than some threshold value} the use of $\kappa-$distributions for the halo population is advantageous, because of drastically larger growth rates compared to those provided by Maxwell halo population. The larger growth rates allow reducing the computation time and make feasible PIC simulations, \blue{which can clarify the efficiency of the electron heat flux regulation by anti-parallel whistler waves}. Second, the presented analysis is valuable for experimental analysis of whistler waves in the solar wind. Coherent whistler waves measured in the solar wind \blue{are usually identified} in the magnetic field spectra as local bumps superimposed on \blue{a spectrum of turbulent magnetic field fluctuations} \cite{Lacombe14,Tong19b:apj}. \blue{The presented analysis shows} that the detection of anti-parallel whistler waves using that methodology is expected to be more complicated than for parallel whistler waves, because anti-parallel whistler waves are expected to have smaller amplitudes and lower frequencies, while the turbulence intensity is higher at lower frequencies. Thus, compared to parallel whistler waves, power spectral density corresponding to coherent anti-parallel whistler waves is more likely to be obscured by turbulent magnetic field fluctuations. An experimental analysis should take into account that feature of whistler waves in the solar wind, otherwise the selection procedure of whistler waves will be biased toward parallel whistler waves. 

{\bf Acknowledgments:}
The work of I.V. and S.B. was supported by NASA grant 80NSSC18K0646. The work of I.K. was supported by the NSF Grant No. 1502923 and the NASA Van Allen Probes RBSPICE instrument project provided by JHU/APL subcontract 131803 under NASA prime contract NNN06AA01C. The work of A.A. was supported by the Russian Science Foundation grant No 19-12-00313. I.V. also thanks for support the International Space Science Institute (ISSI), Bern, Switzerland. I.V. thanks Yuguang Tong, Rachel Wang, Trevor Bowen and Lynn Wilson for discussions. All the data presented in the figures in the manuscript are available from the corresponding author upon reasonable request.

\bibliographystyle{unsrt}


\end{document}